\documentclass[preprint,10pt]{elsarticle}

\usepackage{amssymb}
\usepackage{amsthm}
\usepackage{xcolor} 
\usepackage{amssymb} 
\usepackage{amsmath}
\usepackage{graphicx}

\usepackage{tikz}

\begin{document}

\begin{frontmatter}



\title{Semi-Mamba-UNet: Pixel-Level Contrastive and Pixel-Level Cross-Supervised Visual Mamba-based UNet for Semi-Supervised Medical Image Segmentation}


\author[inst1]{Chao Ma}
\affiliation[inst1]{organization={School of Computer Science and Technology, Xinjiang University},
            city={Ürümqi},
            country={China}}

\author[inst2]{Ziyang Wang\corref{cor1}}
\ead{ziyang.wang17@gmail.com}
\cortext[cor1]{Corresponding author.}
\affiliation[inst2]{organization={Department of Computer Science},
            addressline={University of Oxford}, 
            city={Oxford},
            country={UK}}

\begin{abstract}

Medical image segmentation is essential in diagnostics, treatment planning, and healthcare, with deep learning offering promising advancements. Notably, the convolutional neural network (CNN) excels in capturing local image features, whereas the Vision Transformer (ViT) adeptly models long-range dependencies through multi-head self-attention mechanisms. Despite their strengths, both the CNN and ViT face challenges in efficiently processing long-range dependencies in medical images, often requiring substantial computational resources. This issue, combined with the high cost and limited availability of expert annotations, poses significant obstacles to achieving precise segmentation. To address these challenges, this study introduces Semi-Mamba-UNet, which integrates a purely visual Mamba-based U-shaped encoder–decoder architecture with a conventional CNN-based UNet into a semi-supervised learning (SSL) framework. This innovative SSL approach leverages both networks to generate pseudo-labels and cross-supervise one another at the pixel level simultaneously, drawing inspiration from consistency regularisation techniques. Furthermore, we introduce a self-supervised pixel-level contrastive learning strategy that employs a pair of projectors to enhance the feature learning capabilities further, especially on unlabelled data. Semi-Mamba-UNet was comprehensively evaluated on two publicly available segmentation dataset and compared with seven other SSL frameworks with both CNN- or ViT-based UNet as the backbone network, highlighting the superior performance of the proposed method. The source code of Semi-Mamba-Unet, all baseline SSL frameworks, the CNN- and ViT-based networks, and the two corresponding datasets are made publicly accessible at \textcolor{red}{https://github.com/ziyangwang007/Mamba-UNet}.

\end{abstract}




\end{frontmatter}
\section{Introduction}

Medical image segmentation is essential for enabling precise diagnostics and effective treatment strategies, and deep-learning-based networks, particularly those based on the convolutional neural network (CNN) UNet architecture, have been investigated extensively \cite{ronneberger2015u,milletari2016v,ibtehaz2020multiresunet,wang2021quadruple}. The UNet architecture has a symmetrical encoder-decoder configuration and skip connections at each level. The encoder compresses the input feature map to extract abstract features, which the decoder then uses to reconstruct the image, thereby enhancing the semantic segmentation accuracy. Skip connections are designed to copy and paste features, thereby retaining crucial spatial information and further contributing to the efficacy of the network. UNet has catalysed the development of numerous enhancements. For example, U-Net++ \cite{zhou2018unet++} includes a nested UNet structure with deep supervision mechanisms, whereas Attention UNet \cite{oktay2018attention} incorporates attention gates to bolster the feature-learning capabilities of the decoders. Moreover, Res-UNet \cite{diakogiannis2020resunet} integrates residual learning \cite{he2016deep} into its network blocks. Typically, these UNet modifications aim to leverage advanced network constructs such as DenseNet \cite{huang2017densely}, MobileNet \cite{howard2017mobilenets}, and attention mechanism\cite{woo2018cbam} with UNet to improve the feature learning of CNNs, thereby addressing the intricate challenges associated with segmenting complex anatomical structures such as computed tomography and magnetic resonance imaging (MRI) \cite{zhang2020saunet,wang2021rar,chaurasia2017linknet,li2018h}.

\begin{figure}
\centering  
\includegraphics[width=\linewidth]{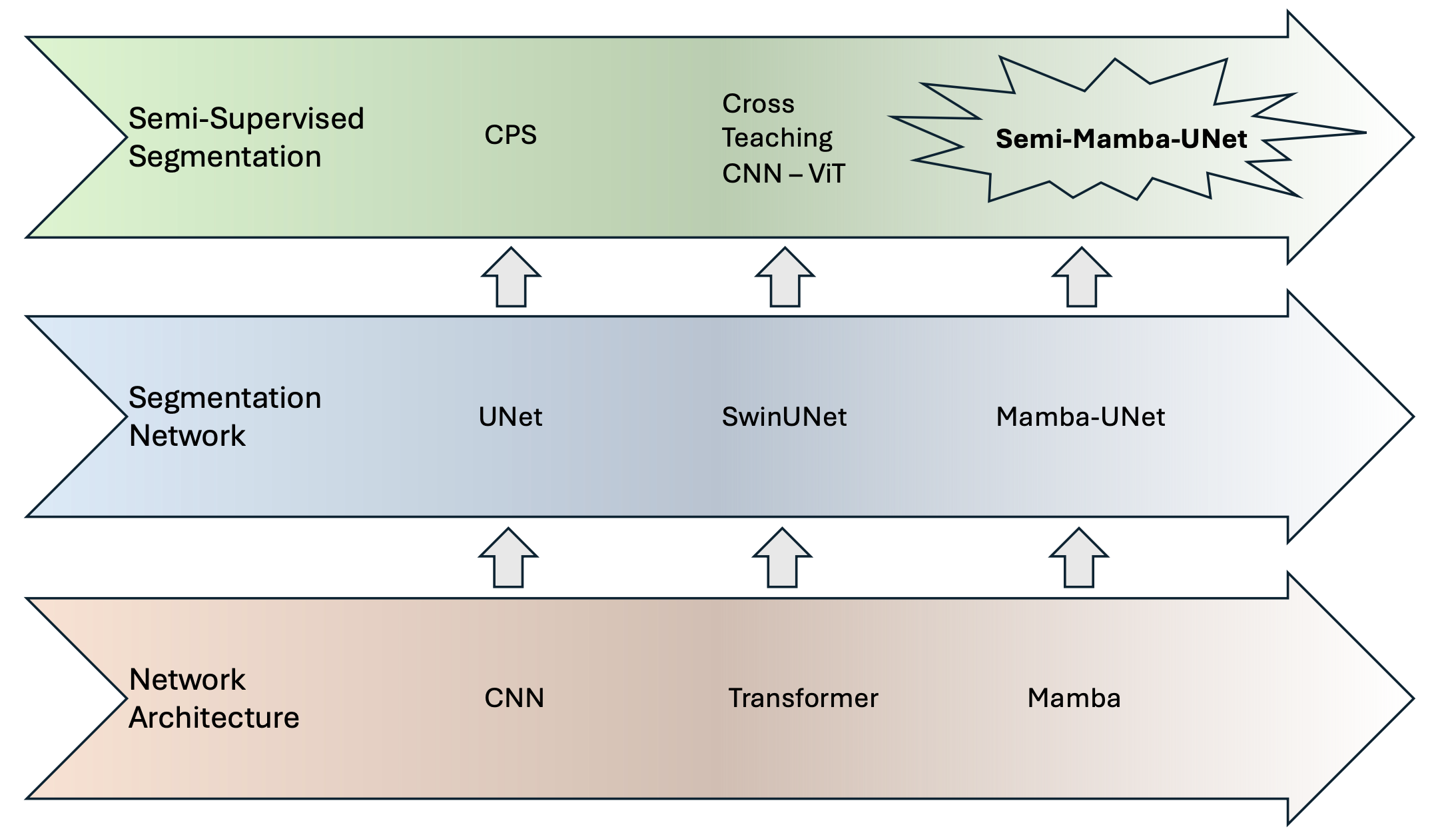}  
\caption{Development history of semi-supervised learning, supervised learning for medical image segmentation, and network architecture. Source: CNN \cite{long2015fully}, Transformer\cite{liu2021swin}, Mamba\cite{liu2024vmamba}, UNet\cite{ronneberger2015u}, Swin-UNet\cite{cao2022swin}, Mamba-UNet\cite{wang2024mamba}, CPS\cite{chen2021semi}, cross-teaching CNN \& ViT\cite{luo2021semi}, and proposed Semi-Mamba-UNet.}
\label{fig:intro}  
\end{figure} 

A recent study on multi-head self-attention in sequence-to-sequence tasks demonstrated the effectiveness of the Transformer network architecture \cite{vaswani2017attention}. Image recognition has been proven to benefit from the Vision Transformer (ViT), which outperforms CNN-based networks, particularly on large datasets, owing to its ability to model long-range dependencies \cite{dosovitskiy2020image}. Several ViT-based networks have been investigated for image segmentation, including SegFormer \cite{xie2021segformer}, Segmenter \cite{strudel2021segmenter}, and SETR \cite{zheng2021rethinking}. In medical image segmentation, most studies relating to ViTs were inspired by UNet, such as TransUNet \cite{chen2021transunet}. This approach explores how Transformer encoders process tokenised image patches from a CNN feature map as the input sequence to extract global contexts, whereas the decoder upsamples the encoded features. These are then combined with high-resolution CNN feature maps to enable precise localisation. Swin-UNet \cite{cao2022swin} further explores the integration of a pure shift window-based ViT into a U-shaped architecture, resulting in a pure Swin ViT-based UNet. Dense Swin-UNet \cite{wang2023densely} advances this by incorporating deep supervision and densely connected skip connections to enhance the segmentation performance. UTNet \cite{gao2021utnet} incorporates a computationally efficient self-attention mechanism along with relative position encoding to reduce the complexity of self-attention operations. UNETR \cite{hatamizadeh2022unetr} includes a ViT-based UNet for volumetric medical image segmentation. nnFormer \cite{zhou2023nnformer}, a 3D Transformer for volumetric medical image segmentation, not only exploits the combination of interleaved convolution and self-attention operations, but also introduces a local and global volume-based self-attention mechanism to learn the volume representations.

The efficacy of ViT-based networks, while promising, is contingent on the availability of extensively labelled datasets, which are challenging to acquire. Weakly supervised learning (WSL) and semi-supervised learning (SSL) frameworks have been investigated \cite{luo2022scribble,wang2023exigent,yu2019uncertainty,li2020transformation,wang2023weakly}. A common approach in these studies involves the integration of UNet with consistency regularisation strategies, wherein the network is encouraged to produce consistent outputs under various perturbations. For example, the Uncertainty-Aware Mean Teacher (UAMT) method employs the UNet architecture within a self-ensembling scheme for feature perturbation and uncertainty estimation \cite{yu2019uncertainty}. The cross-teaching technique extends this concept by leveraging CNN- and ViT-based UNet, enabling collaboration between the two networks through pseudo-labels\cite{luo2021semi}. FixMatch introduces a novel approach that employs both strong and weak data augmentation as forms of data perturbation across networks \cite{sohn2020fixmatch}. Furthermore, multi-view learning expands this cooperative framework to include three networks, thereby promoting mutual learning through co-training \cite{xia20203d}.

Recent advancements have resulted in the novel Mamba architecture, with strength in capturing global contextual information with efficient computational costs, conceptualised by a state-space model (SSM) \cite{wang2023selective,gu2023mamba,gu2023modeling}. This architecture has been explored in various computer vision tasks, such as Vision Mamba \cite{zhu2024vision}, UMamba \cite{ma2024u}, Segmamba \cite{xing2024segmamba}, MambaUNet \cite{wang2024mamba}, VM-UNet \cite{ruan2024vm}, and Weak-Mamba-UNet \cite{wang2024weak}. In response to the growing need for efficient medical image segmentation, particularly in SSL with limited annotations, this study introduces Semi-Mamba-UNet, which is a novel framework that integrates the Mamba architecture within pixel-level contrastive and cross-supervised learning for semi-supervised medical image segmentation. 

To the best of our knowledge, this is the first study to explore the Mamba architecture in medical image segmentation with limited annotations. The development history of UNet and its derivatives in medical image segmentation, and the position of Semi-Mamba-UNet, is depicted in Figure \ref{fig:intro}. Our contributions are fivefold:

\begin{enumerate}
    \item Recent advancements of Visual Mamba \cite{liu2024vmamba} as a network block in a U-shaped encoder-decoder style network are explored for medical image segmentation.
    \item A Mamba-based segmentation network is integrated with SSL, providing a large amount of unlabelled data for network training. For fair evaluation, comparisons are drawn against the CNN-based UNet \cite{ronneberger2015u} and ViT-based Swin-UNet \cite{cao2022swin} across various SSL frameworks.
    \item A pixel-level contrastive learning strategy is introduced with SSL, incorporating a pair of projectors to maximise the feature learning capabilities using both labelled and unlabelled data.
    \item Pixel-level cross-supervised learning is introduced with the SSL. The network is trained with the help of the other network via pseudo-labelling, thereby extending the utility of unlabelled data in network training.
    \item Semi-Mamba-UNet is validated using two public benchmark datasets, demonstrating state-of-the-art performance. The source code of Semi-Mamba-UNet and all baseline methods are made publicly available.
\end{enumerate}

\begin{figure*}[htbp]
\centering  
\includegraphics[width=\linewidth]{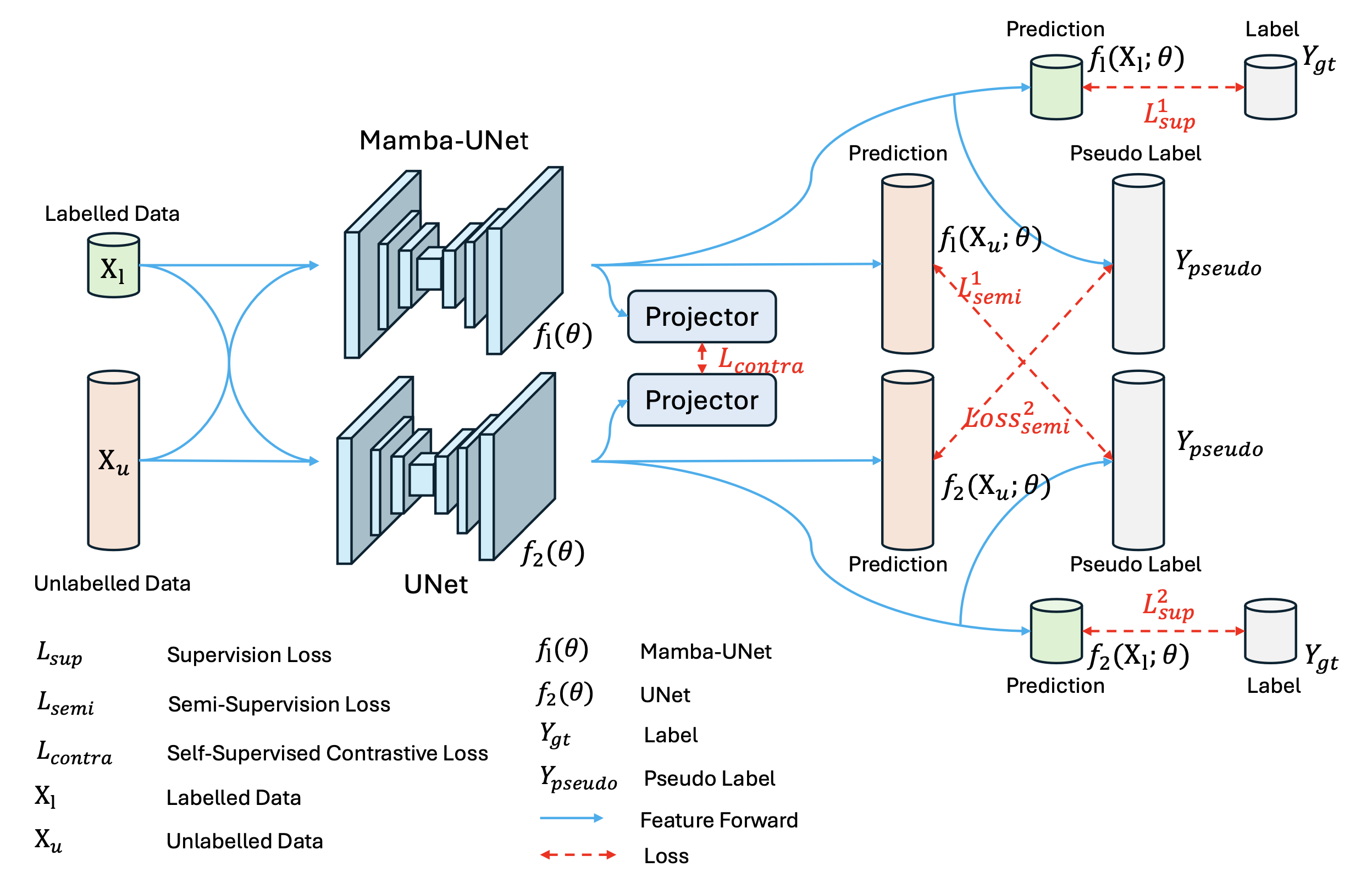}  
\caption{Semi-Mamba-UNet: Framework for pixel-level contrastive cross-supervised Visual Mamba-based UNet for semi-supervised medical image segmentation.}
\label{fig:framework}  
\end{figure*}

\section{Related Work}

\subsection{Mamba in Medical Image Segmentation}

Since the introduction of fully convolutional networks for segmentation in 2015, CNNs have proven to be highly efficient in capturing local feature information in medical image segmentation \cite{long2015fully}. The UNet architecture, with its symmetrical encoder-decoder structure, has further advanced CNN-based image segmentation techniques, resulting in various CNN-based approaches \cite{ronneberger2015u,zhou2018unet++,oktay2018attention,zhang2020saunet,wang2021rar,chaurasia2017linknet,li2018h,cciccek20163d,ma2024triconvunext,isensee2018nnu}. With the emergence of the Transformer architecture \cite{vaswani2017attention}, which was initially designed for sequence-to-sequence tasks, researchers have begun to explore its potential for medical image segmentation owing to its ability to model long-range dependencies, as demonstrated by ViT \cite{dosovitskiy2020image}. Various ViT-based and hybrid CNN-ViT networks have been investigated \cite{chen2021transunet,hatamizadeh2022unetr,cao2022swin,zhou2023nnformer,gao2021utnet,wang2023densely}.

In recent years, the introduction of the Mamba architecture has been aimed at enhancing the efficiency compared with the CNN and Transformer. The original Mamba block, which integrates a gated MLP into the SSM of H3 \cite{fu2022hungry}, uses an SSM that is sandwiched between two gated connections alongside a standard local convolution with SiLU \cite{hendrycks2016gaussian} or Swish \cite{ramachandran2017swish} activation functions. The Mamba architecture is characterised by a repetition of Mamba blocks interleaved with standard normalisation and residual connections \cite{he2016deep}. Although originally designed for sequence tasks, adaptations such as Bidirectional Mamba and Visual State Space have been developed for vision-related tasks \cite{zhu2024vision}. Other Mamba-related backbones include LocalMamba \cite{huang2024localmamba}, PlainMamba \cite{yang2024plainmamba}, MambaMixer \cite{behrouz2024mambamixer}, and Simba \cite{patro2024simba}. Mamba has shown significant potential for 2D medical image segmentation, which is often inspired by UNet, leading to variants such as Mamba-UNet \cite{wang2024mamba}, H-VMUNet \cite{wu2024h}, U-Mamba \cite{ma2024u}, and VM-UNet \cite{ruan2024vm}. Other non-UNet style approaches for 2D segmentation include ProMamba \cite{xie2024promamba} and PMamba \cite{ye2024p}. Several methods have been proposed for 3D medical image segmentation, including Lightm-UNet \cite{liao2024lightm}, SegMamba \cite{xing2024segmamba}, and T-Mamba \cite{hao2024t}, which enable more accurate and comprehensive diagnoses. An increasing number of Mamba-based segmentation networks have been developed, motivated by past CNN and ViT studies, and have explored different modalities of medical images. The training of such advanced networks with limited annotations requires further exploration.

\subsection{Medical Image Segmentation with Limited Annotations}

Despite the promising performance of CNN-, ViT-, and Mamba-based segmentation networks, these networks typically require large, well-annotated datasets that are often difficult to obtain owing to high annotation costs. To address this issue, SSL and self-supervised learning strategies have been employed to leverage limited annotations with large amounts of raw data. Consistency regularisation in SSL ensures the consistency of inferences under various perturbations \cite{tarvainen2017mean,verma2019interpolation,yu2019uncertainty,french2019semi}. Some studies have applied perturbations to input images, augmenting them randomly and setting consistency constraints among the inferences of these augmented images \cite{french2019semi,kim2020structured}. Perturbations can also be applied to feature information, such as in feature perturbation schemes in which multiple decoders are used, and the differences in their inferences are minimised through cross-consistency training \cite{ouali2020semi}.

A common SSL framework involves a student–teacher model \cite{laine2016temporal,tarvainen2017mean,wang2022uncertainty}, where the student model is trained with labelled data and perturbations and the teacher model parameters are updated based on the average weights of the student. This makes the teacher model more robust by guiding students with pseudo-labels under consistency-aware constraints \cite{yu2019uncertainty}. In addition to consistency regularisation, adversarial learning techniques employ an additional discriminator model to distinguish between ground-truth segmentations and model inferences \cite{mittal2019semi,hung2018adversarial}. In this iterative process, the segmentation model and discriminator are trained against one another, with the discriminator aiming to validate high-quality inferences as pseudo-labels and the segmentation model striving to produce confident predictions against the discriminator \cite{zhang2017deep,hung2018adversarial,miyato2018virtual}.

Self-supervised learning, particularly contrastive learning, is another approach for enhancing the training performance. Contrastive learning aims to learn an embedding space in which similar images are closer together and dissimilar images are pushed apart, and has been successful in various tasks, including feature representation learning \cite{chen2020simple,he2020momentum} and unsupervised domain adaptation \cite{kang2019contrastive}. In medical image analysis, contrastive learning has been used to address challenges associated with limited annotations and improve the feature learning capabilities, thereby enhancing the overall performance \cite{chaitanya2020contrastive,hu2021semi,you2022simcvd}. Researchers continue to explore different strategies for contrastive learning, such as augmentation techniques, similarity metrics, and negative sample mining, to optimise the effectiveness of these networks further \cite{wu2022cross,lou2023min}. In this study, we propose leveraging SSL and self-supervised learning simultaneously for Mamba-based networks with limited annotations.

\section{Methodology}

The framework of Semi-Mamba-UNet is illustrated in Figure~\ref{fig:framework}. As shown in the figure, $(\textbf{\textit{X}}_{\rm l}, \textbf{\textit{Y}}_{\rm gt}) \in \mathbf{L}$ denotes the labelled training dataset, whereas $(\textbf{\textit{X}}_{\rm u}) \in \mathbf{U}$ denotes the unlabelled training dataset. In addition, $(\textbf{\textit{X}}_{\rm t}, \textbf{\textit{Y}}_{\rm t}) \in \mathbf{T}$ denotes the labelled testing dataset. $\textbf{\textit{X}} \in \mathbb{R}^{h \times w}$ represents a 2D grayscale image of size $h$ height and $w$ width. $\textbf{\textit{Y}}_{\rm l}, \textbf{\textit{Y}}_{\rm t} \in \mathbb{N}_{4}^{h \times w}$ represents a four-class labelled segmentation mask with pixel values ranging from 0 to 3. The segmentation mask predicted by a segmentation network given $\textbf{\textit{X}}$ is $\textit{Y}_{\rm p} = f(\textit{X};\theta)$ with $\theta$ as the parameters. The Mamba-based UNet and UNet are denoted by $f_{1}(\theta)$ and $f_{2}(\theta)$, respectively. The prediction of a network can be considered as a pseudo-label to expand the unlabelled dataset to $(\textbf{\textit{X}}_{\rm u}, \textbf{\textit{Y}}_{\rm pseudo}) \in \mathbf{U}$ to train the other network. A pair of projectors $p(\cdot)$ is introduced into each network to extract the representation features of the training set for contrastive learning. The overall losses are categorised as the supervision loss $\mathcal{L}_{\rm sup}$, semi-supervised loss $\mathcal{L}_{\rm semi}$, and self-supervised contrastive loss $\mathcal{L}_{\rm contra}$. The evaluation is conducted by measuring the difference between $(\textbf{\textit{Y}}_{\rm p}, \textbf{\textit{Y}}_{\rm t})$ in the test set. The overall training objective is to update the network parameters $\theta$ thus minimising the total loss $\mathcal{L}_{total}$, which is expressed as
\begin{equation}
\label{loss}
 \mathcal{L}_{total} = \underbrace{ \mathcal{L}_{\text{sup}}^{1} + \mathcal{L}_{\text{sup}}^{2}}_{sup} + \underbrace{\mathcal{L}_{\text{semi}}^{1} + \mathcal{L}_{\text{semi}}^{2}}_{semi} + \underbrace{\mathcal{L}_{\text{contra}}}_{self}
\end{equation}

All mathematical symbols are denoted in Figure~\ref{fig:framework}, and $\mathcal{L}$ is highlighted by a red dashed line, where $sup$ is the supervision loss for $f_{1}(\theta)$ and $f_{2}(\theta)$ based on the labelled training set. $\mathcal{L}_{\text{sup}}$ is designed with a combination of the Dice coefficient-based (Dice) and cross-entropy-based (CE) losses, as follows:

\begin{equation}
\mathcal{L}_{\rm sup}^{1} = \mathrm{CE}\big({\rm softmax}{(f_{1}( {\textbf{\textit{X}}_{u}}; \theta),  \textbf{\textit{Y}}_{\rm gt})}  \big) + \mathrm{Dice}\big({\rm softmax}{(f_{1}( {\textbf{\textit{X}}_{u}}; \theta), \textbf{\textit{Y}}_{\rm gt} )}  \big)
\end{equation}
\begin{equation}
\mathcal{L}_{\rm sup}^{2} = \mathrm{CE}\big({\rm softmax}{(f_{2}( {\textbf{\textit{X}}_{u}}; \theta), \textbf{\textit{Y}}_{\rm gt} )}  \big) + \mathrm{Dice}\big({\rm softmax}{(f_{2}( {\textbf{\textit{X}}_{u}}; \theta), \textbf{\textit{Y}}_{\rm gt} )}  \big)
\end{equation}

$\mathcal{L}_{\text{semi}}$ is the semi-supervision loss for $f_{1}(\theta)$ and $f_{2}(\theta)$ based on the unlabelled training set. The prediction of a network is considered as the pseudo-label $\textbf{\textit{Y}}_{\rm pseudo}$ to extend $\textbf{\textit{X}}_{\rm u}$ to retrain the other networks. $\mathcal{L}_{\text{contra}}$ is the self-supervised contrastive learning loss, and we propose a projector pair to extract the features between the predictions of the two networks. The details of Mamba-UNet, pixel-level cross-supervised learning with $\mathcal{L}_{\rm semi}$, and pixel-level contrastive learning with $\mathcal{L}_{\rm contra}$ are discussed in the following sections.

\subsection{Mamba-UNet}

\begin{figure*}
\centering  
\includegraphics[width=\linewidth]{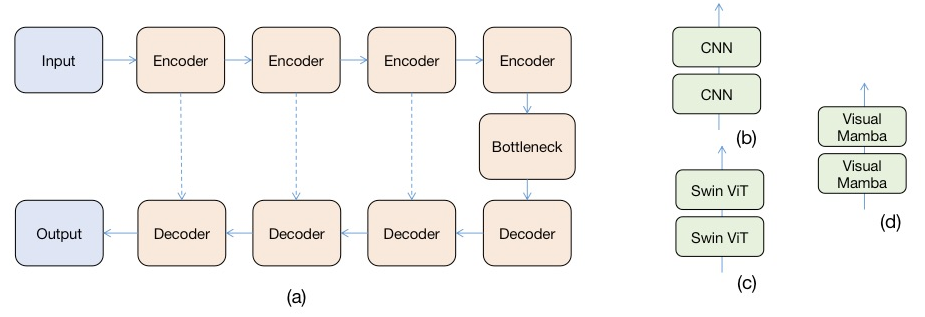}  
\caption{Segmentation backbone network in this study. (a) Encoder-decoder style segmentation network. (b) Two-layer CNN-based network block of UNet. (c) Two-layer Swin ViT-based network of Swin-UNet. (d) Two-layer Visual Mamba-based network block of Mamba-UNet.}
\label{fig:network}  
\end{figure*}

The UNet architecture, as depicted in Figure~\ref{fig:network}, represents a novel adaptation of the conventional encoder-decoder style segmentation network with various types of network blocks for medical image analysis. To ensure a fair comparison, the proposed utilisation of Mamba-UNet is developed against the original UNet \cite{ronneberger2015u} and Swin-UNet \cite{cao2022swin}. Each of these networks adheres to the U-shaped encoder-decoder configuration. Specifically, UNet employs a two-layer CNN with a size of $3 \times 3$\cite{ronneberger2015u}, Swin-UNet utilises two Swin Transformer blocks \cite{cao2022swin}, and Mamba-UNet integrates two Visual Mamba blocks, which is part of our previous work in \cite{wang2024mamba}. This distinction in the block composition is pivotal, because it directly influences the ability of the network to process and interpret the intricate details present in medical images. 

Specifically, conventional SSMs are used as a linear time-invariant system function to map $x(t) \in \mathbb{R} \mapsto y(t) \in \mathbb{R}$ through a hidden state $h(t) \in \mathbb{R}^N$, given $A \in \mathbb{C}^{N \times N}$ as the evolution parameter, $B, C \in \mathbb{C}^{N}$ as the projection parameters for a state size $N$, and the skip connection $D \in \mathbb{C}^{1}$. The model can be formulated as the linear ordinary differential equations in Eq. \ref{ode}:

\begin{equation}
\label{ode}
\begin{aligned}
h'(t) &= Ah(t) + Bx(t), \\
y(t) &= Ch(t) + Dx(t).
\end{aligned}
\end{equation}

The discrete version of this linear model can be transformed by a zero-order hold, given a timescale parameter $\Delta \in \mathbb{R}^{D}$.

\begin{equation}
\begin{aligned}
h_t &= \overline{A}h_{k-2} + \overline{B}x_k \\
y_t &= {C}h_k + \overline{D}x_k\\
\overline{A} &= e^{\Delta A} \\
\overline{B} &= (e^{\Delta A} - I) A^{-1}B \\
\overline{C} &= C 
\end{aligned}
\end{equation}
where $B,C \in \mathbb{R}^{D \times N}$. The approximation of $\overline{B}$ is refined using the first-order Taylor series $\overline{B} = \left(e^{\Delta A} - I\right) A^{-1} B \approx \left(\Delta A\right)\left(\Delta A\right)^{-1} \Delta B = \Delta B$. Visual Mamba further introduces the cross-scan module and integrates convolutional operations into the block, as detailed in \cite{gu2023mamba}. Mamba-UNet, with its Visual Mamba blocks, aims to capitalise on the efficiency and effectiveness of Mamba models in capturing and processing complex spatial and contextual information, thereby enhancing the segmentation performance.

\subsection{Pixel-Level Cross-Supervised Learning}

Inspired by the principles of consistency regularisation and multi-view learning, such as cross-pseudo-supervision \cite{chen2021semi}, where two independently initialised networks generate and exchange pseudo-labels for mutual supervision, this study extends the concept to leverage the complementary strengths of distinct architectures. The methodology of cross-teaching between a CNN and ViT \cite{luo2022semi} further explores the mutual benefits derived from the collaboration between two different network architectures. Similarly, FixMatch \cite{sohn2020fixmatch} advocates for the application of two distinct data augmentations across two networks, with one network acting as a supervisor for the other using augmented data. In Semi-Mamba-UNet, we introduce a simple yet efficient cross-supervised learning strategy that enables Mamba-UNet and UNet to help each other directly; $\mathcal{L}_{\text{semi}}$ is illustrated as
\begin{equation}
\mathcal{L}_{\rm semi}^{1} = \mathrm{CE}\big({\rm argmax}{(f_{1}( {\textbf{\textit{X}}_{u}}; \theta), f_{2}( {\textbf{\textit{X}}_{u}}; \theta) )}  \big) + \mathrm{Dice}\big({\rm argmax}{(f_{1}( {\textbf{\textit{X}}_{u}}; \theta), f_{2}( {\textbf{\textit{X}}_{u}}; \theta) )}  \big)
\end{equation}
\begin{equation}
\mathcal{L}_{\rm semi}^{2} = \mathrm{CE}\big({\rm argmax}{(f_{2}( {\textbf{\textit{X}}_{u}}; \theta), f_{1}( {\textbf{\textit{X}}_{u}}; \theta) )}  \big) + \mathrm{Dice}\big({\rm argmax}{(f_{2}( {\textbf{\textit{X}}_{u}}; \theta), f_{1}( {\textbf{\textit{X}}_{u}}; \theta) )}  \big)
\end{equation}

\subsection{Pixel-Level Contrastive Learning}
Contrastive learning has been recognised as a potent paradigm for deriving robust and discriminative features, representing a significant stride in the realm of self-supervised learning \cite{oord2018representation}. The key concept underlying contrastive learning is that the positive and negative samples are discriminative. Positive and negative samples are initially constructed based on prior knowledge and mapped onto a potential feature embedding space. A metric function is then utilised to encourage the network to bring positive samples closer together and distance the positive samples from the negative samples. Contrastive learning has exhibited significant efficacy in various applications \cite{chen2020simple,he2020momentum,kang2019contrastive}.

The application of contrastive learning in medical image analysis addresses the perennial challenges posed by sparse annotations, and augments the capacity for feature extraction, culminating in enhanced model performance \cite{chaitanya2020contrastive,hu2021semi,you2022simcvd,wang2023dual,shi2023aging}. Most of the latest methods construct positive and negative samples using specific tracks, such as applying different perturbations to the same sample and encoding the same sample with different types of encoders,  focusing on the variability of the sample attributes. To understand the variability of pixels between samples better, a novel pixel-level contrastive learning approach is proposed.

Considering the small size of cardiac data, while a large amount of pixels belongs to the background, the consistency of these pixels is insignificant for network training. Adaptive average pooling is used as a projector to filter unwanted background pixels and highlight the representational ability of the target region in the image. Furthermore, we apply L2 regularisation in the channel dimension to sparsify the features and improve the resistance of the model to perturbation. Inter-network consistency is also used by computing the mean square error between features.
In the proposed SSL framework, we utilise a projector pair to Mamba-UNet and UNet simultaneously. This configuration facilitates the extraction of pixel-level feature representations, which subsequently serve as the basis for computing the image similarity within the defined feature space. A similarity assessment is conducted according to \cite{xie2020pgl}, which is formalised as follows:

\begin{equation}
\mathcal{L}_{\rm contra} = \frac{\sum \left\| \left(G(F_{\theta}(X_L  \cup X_U)), G(F_{\theta}(X_L  \cup  X_U))\right) \right\|_2^2 }{N}
\end{equation}
where $F_{\theta}$ is a predictor with the same AdaptiveAvgPool as the projector, $G$ is $l_2$ regularisation along the channel axis, and $N$ is the number of input data points. $X_L$ and $X_U$ represent labelled and unlabelled data, respectively, and $\cup$ represents union with a mathematical symbol. To leverage the dataset for network training effectively, we further assume labelled data as unlabelled data to expand the dataset (i.e. $\cup$) in the process of consistency regularisation of the unlabelled set to boost the performance, which differs from conventional SSL strategies.

\section{Experiments and Results}

\textbf{Datasets:} The efficacy of Semi-Mamba-UNet, alongside various baseline methodologies, was assessed using two publicly available datasets. {\bf ACDC} is an MRI cardiac segmentation dataset from the MICCAI 2017 Challenge \cite{bernard2018deep}. This dataset encompasses imaging data from 100 cases and provides a comprehensive basis for evaluation. ACDC was designed for a four-class segmentation task with three regions of interest: the right ventricle (RVC), left ventricle (LVC), and myocardium (MYO). {\bf PROMISE12} is an MR prostate segmentation dataset obtained from the MICCAI 2012 Challenge \cite{litjens2014evaluation}. This dataset encompasses imaging data from 50 cases. All images are transversal T2-weighted MR images of the prostate. PROMISE12 was designed for binary segmentation tasks. To comply with the input requirements of Swin-UNet \cite{cao2022swin} and MambaUNet \cite{wang2024mamba}, all images were resized to $224 \times 224$ pixels. The dataset was partitioned such that 20\% constituted the testing set, and the remaining 80\% was allocated for training and validation. The experimental setup was designed to simulate scenarios in which only three and seven cases of the training set were available as labelled data for ACDC, with 100 cases provided. For PROMISE12, we assumed that eight and 12 cases were labelled, with 50 cases provided in total. The \&labelled  unlabelled cases were randomly selected from ACDC and PROMISE12 only once and utilised for Semi-Mamba-UNet and all baseline methods. The list of labelled training, unlabelled training, validation, and testing sets can be found on the GitHub page, and there was no overlap among the subsets.

\begin{figure*}
\centering  
\includegraphics[width=\linewidth]{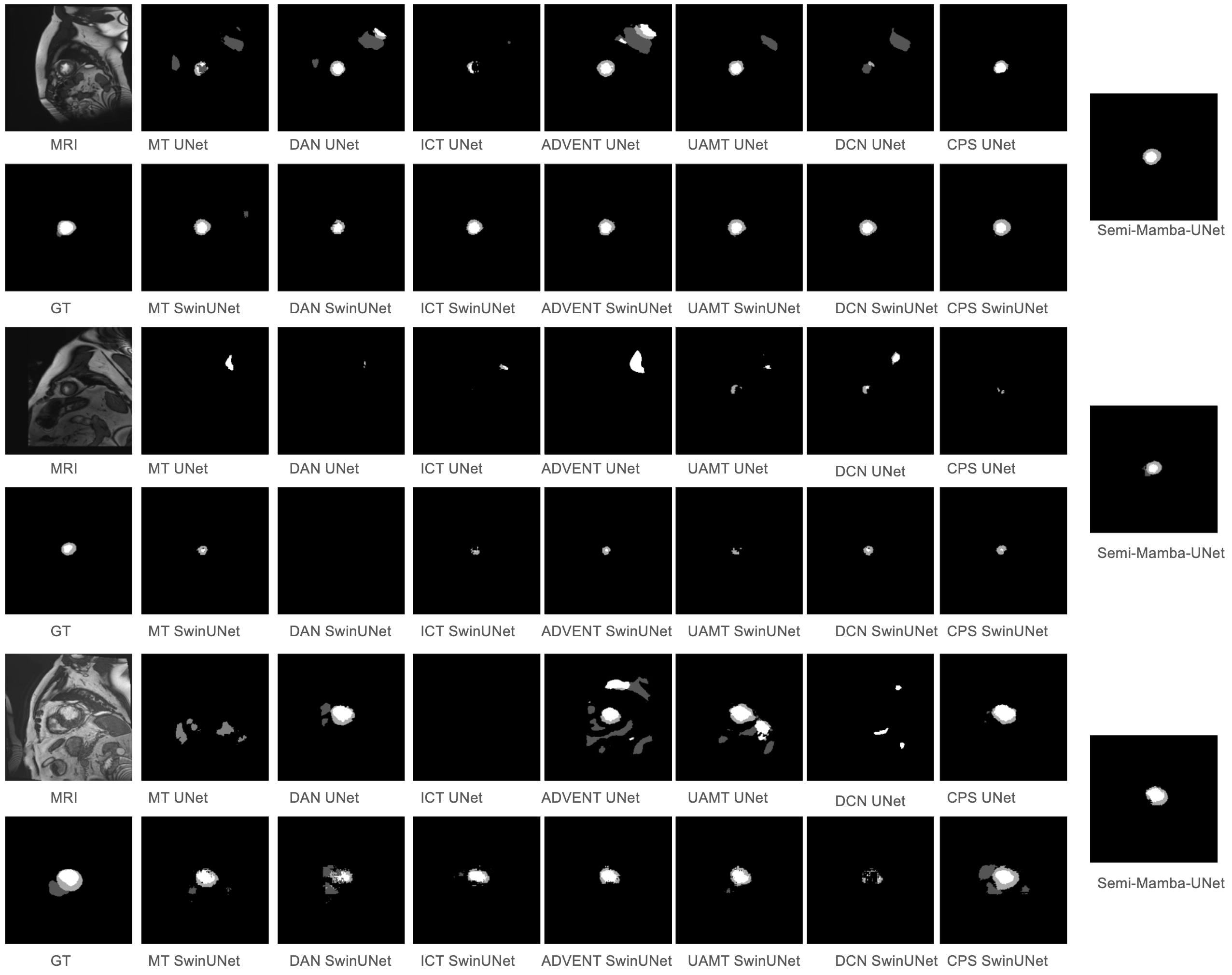}  
\caption{Three randomly selected example MRI images in MRI cardiac test set, ground truth, and corresponding segmentation results of all baseline methods and Semi-Mamba-UNet when three cases of data were assumed as labelled data.}
\label{fig:3}  
\end{figure*} 

\begin{figure*}
\centering  
\includegraphics[width=\linewidth]{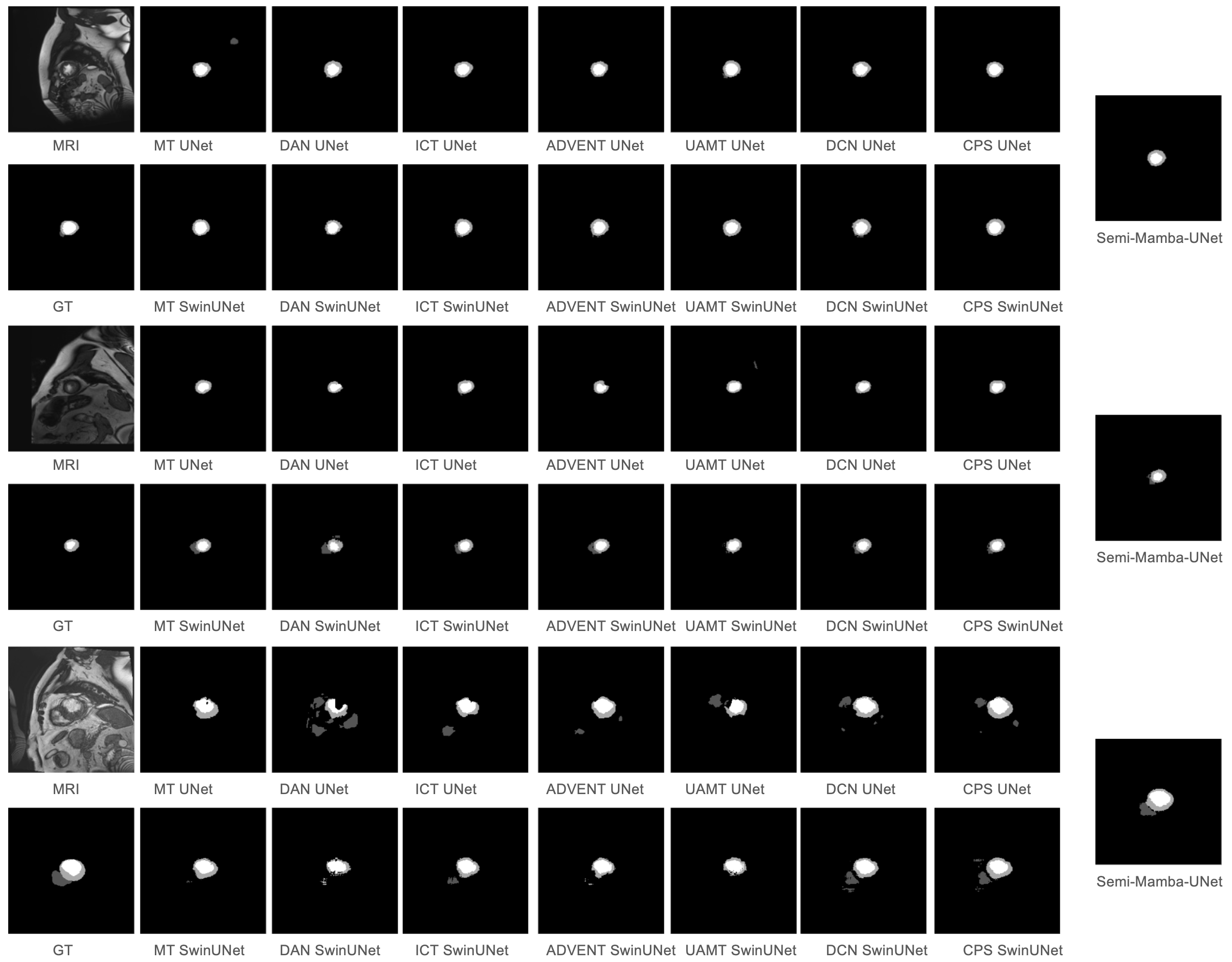}  
\caption{Three randomly selected example MRI images in MRI cardiac test set, ground truth, and corresponding segmentation results of all baseline methods and Semi-Mamba-UNet when five cases of data were assumed as labelled data.}
\label{fig:7}  
\end{figure*}

\begin{figure*}
\centering  
\includegraphics[width=\linewidth]{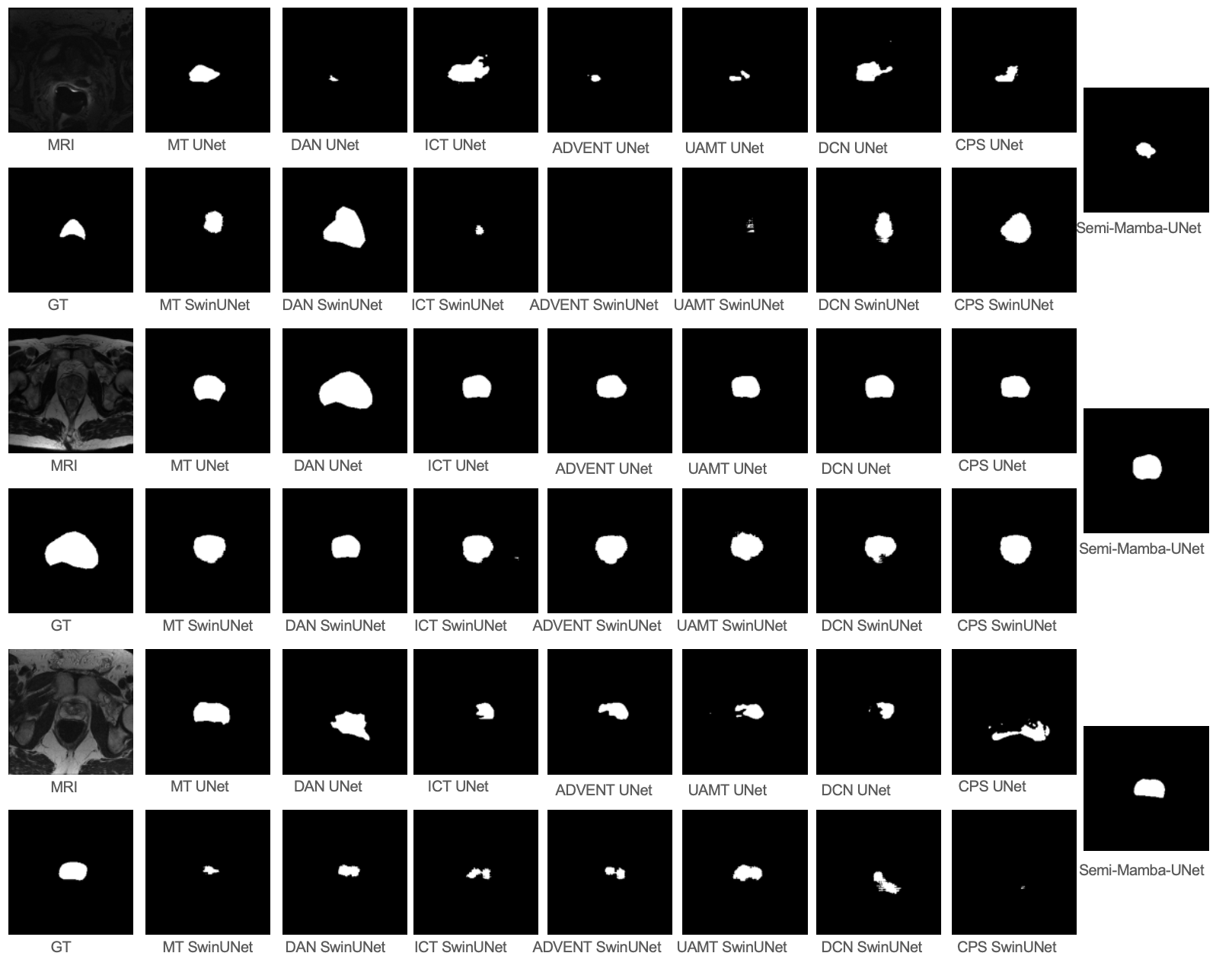}  
\caption{Three randomly selected example MRI images in MRI prostate test set, ground truth, and corresponding segmentation results of all baseline methods and Semi-Mamba-UNet when eight cases of data were assumed as labelled data.}
\label{fig:8}  
\end{figure*} 

\begin{figure*}
\centering  
\includegraphics[width=\linewidth]{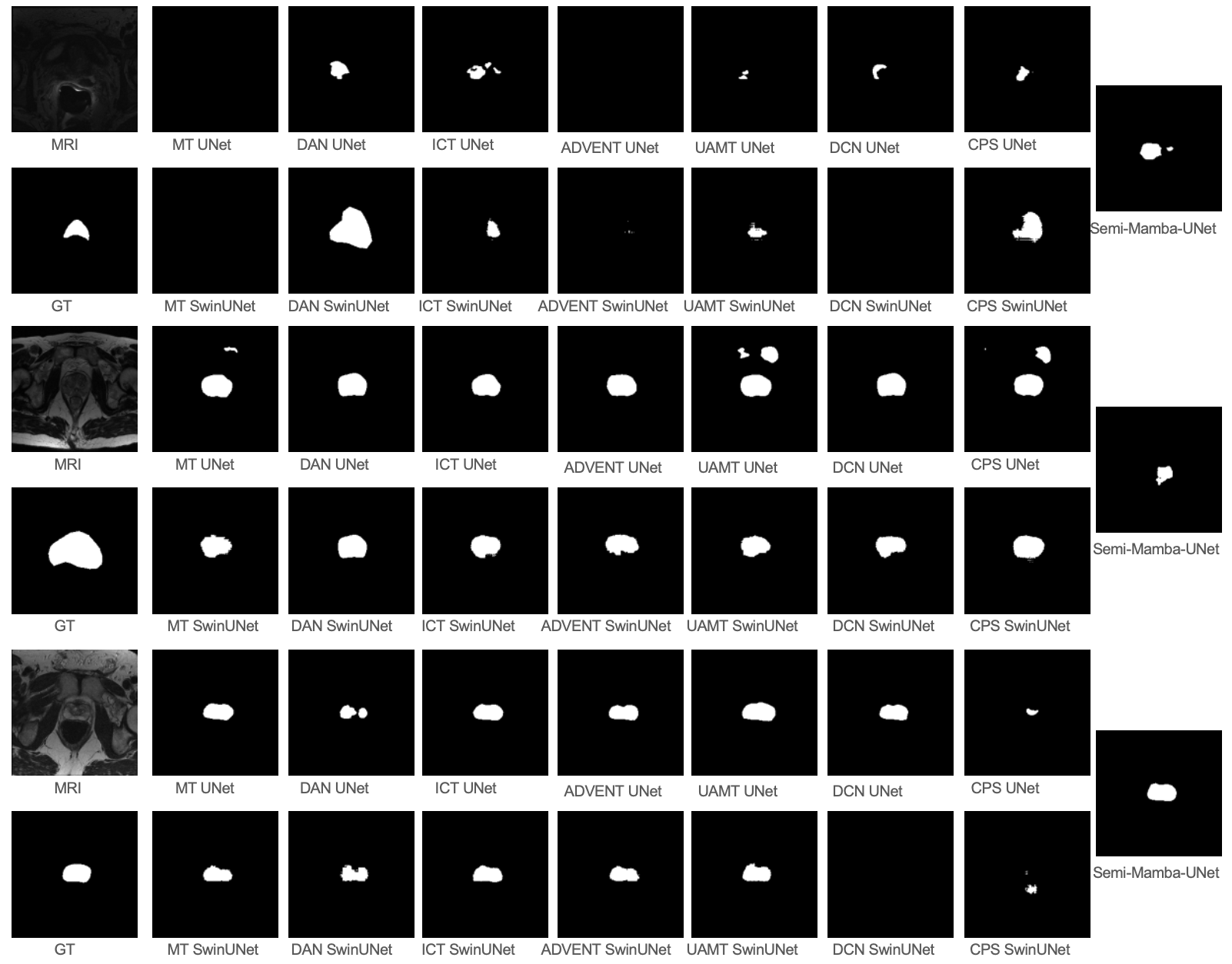}  
\caption{Three randomly selected example MRI images in MRI prostate test set, ground truth, and corresponding segmentation results of all baseline methods and Semi-Mamba-UNet when 12 cases of data were assumed as labelled data.}
\label{fig:12}  
\end{figure*} 

\textbf{Implementation Details:} The development environment for our experiments was Ubuntu 20.04, utilizing PyTorch. The computational hardware included an Nvidia GeForce RTX 3090 GPU and Intel Core i9-10900K CPU. The average runtime for the experiments was approximately 5 to 8 h for ACDC and 2 to 5 h for PROMISE12. Both datasets were designed for 2D image segmentation tasks. The Semi-Mamba-UNet training encompassed 30,000 iterations with a batch size of 16. A stochastic gradient descent optimiser was employed, with a learning rate of 0.01, momentum of 0.9, and weight decay of 0.0001. The validation set was evaluated every 200 iterations, with the network weights preserved only if the validation performance surpassed that of the previous best network.

\textbf{Baseline Segmentation Networks:} The framework of Semi-Mamba-UNet is depicted in Figure~\ref{fig:framework}, with two segmentation backbone networks. To ensure equitable comparisons, we also employed the CNN-based UNet \cite{ronneberger2015u} and ViT-based Swin-UNet \cite{cao2022swin} as the segmentation backbone networks for different SSL frameworks. The total numbers of parameters for each network are listed in Table \ref{tablecost}. This selection was motivated by the architectural similarities that these networks share with our proposed framework, thereby providing a consistent basis for evaluating the performance enhancements introduced by Semi-Mamba-UNet.

\begin{table}[htbp]
\caption{ Computational cost of segmentation backbone networks.}
\centering
\begin{tabular}{c|ccc}
\hline
Network & CNN-based UNet & ViT-based UNet & Mamba-based UNet \\
\hline
Parameters & 1,813,764 & 27,168,420 & 19,121,472\\
\hline
\end{tabular}
\label{tablecost}
\end{table}

\textbf{Baseline SSL Frameworks:} The SSL baseline frameworks evaluated included the Mean Teacher (MT)~\cite{tarvainen2017mean}, deep adversarial network (DAN)~\cite{zhang2017deep}, interpolation consistency training (ICT)~\cite{verma2019interpolation}, Adversarial Entropy Minimization (ADVENT)~\cite{vu2019advent}, UAMT\~cite{yu2019uncertainty}, and deep co-training (DCN)~\cite{qiao2018deep}. Both Swin-UNet~\cite{cao2022swin} and UNet~\cite{ronneberger2015u} were employed as the segmentation backbone networks across all SSL frameworks.

\begin{table*}[htbp]
\footnotesize
\caption{Direct comparison of semi-supervised frameworks on MRI cardiac test set when three cases of 100 data were assumed as labelled data.}
\centering
\begin{tabular}{c|ccccc|cc}
\hline
SSL Framework+Network & Dice$\uparrow$  & Acc$\uparrow$ & Pre$\uparrow$ & Sen$\uparrow$ & Spe$\uparrow$ & HD$\downarrow$ & ASD$\downarrow$  \\
\hline
MT\cite{tarvainen2017mean} + Swin-UNet & 0.7506 & 0.9910 & 0.7918 & 0.7178 &  0.9394 & 10.4621 & 3.5301 \\
DAN\cite{zhang2017deep} + Swin-UNet &  0.7252 & 0.9901 & 0.7695 & 0.6903 & 0.9337& 12.9800 & 4.3823\\
ICT\cite{verma2019interpolation} + Swin-UNet& 0.7504 & 0.9910 & 0.7923 & 0.7180 & 0.9392 &9.8026 & 3.1055  \\
ADVENT\cite{vu2019advent}+ Swin-UNet  & 0.7489 & 0.9910 & 0.7964 & 0.7128 & 0.9373 & 11.1535 & 3.0907\\
UAMT\cite{yu2019uncertainty} + Swin-UNet\cite{wang2022uncertainty} & 0.7442 & 0.9909 & 0.7902 & 0.7108 & 0.9393 & 10.2955 & 2.8222 \\
DCN\cite{qiao2018deep} + Swin-UNet &0.7603 & 0.9914 & 0.8118 & 0.7207 & 0.9376 & 10.1783 & 3.1478 \\
\hline
MT\cite{tarvainen2017mean} + UNet  & 0.7256 & 0.9885 & 0.8217 & 0.6670 & 0.9044 & 24.0480 & 9.7662 \\
DAN\cite{zhang2017deep} + UNet & 0.7657 & 0.9905 & 0.8296 & 0.7152 & 0.9199 & 21.1226 & 7.3434\\
ICT\cite{verma2019interpolation} + UNet  &  0.7490 & 0.9906 & 0.8827 & 0.6633 & 0.9013 & 11.2109 & 4.5181  \\
ADVENT\cite{vu2019advent}+ UNet & 0.6656 & 0.9833 & 0.6487 & 0.6900 & 0.9190 & 42.8011 & 16.6207 \\
UAMT\cite{yu2019uncertainty} + UNet & 0.7472 & 0.9901 & 0.8164 & 0.6943 & 0.9168& 21.7492 & 7.7489 \\
DCN\cite{qiao2018deep} + UNet & 0.7312 & 0.9894 & 0.8316 & 0.6626 & 0.9022 & 24.6607 & 10.1996 \\
\hline
Semi-Mamba-UNet & {\bf0.8386} & {\bf0.9936} &{\bf 0.8861} & {\bf0.7992} & {\bf0.9483} & \underline{6.2139} & \underline{1.6406}  \\
\hline
\end{tabular}
\label{tablebaseline111}
\end{table*}

\begin{table*}[htbp]
\footnotesize
\caption{Direct comparison of semi-supervised frameworks on MRI cardiac test set when seven cases of 100 data were assumed as labelled data.}
\centering
\begin{tabular}{c|ccccc|cc}
\hline
Framework+Network & Dice$\uparrow$  & Acc$\uparrow$ & Pre$\uparrow$ & Sen$\uparrow$ & Spe$\uparrow$ & HD$\downarrow$ & ASD$\downarrow$  \\
\hline
MT\cite{tarvainen2017mean} + Swin-UNet &0.8678 & 0.9949 & 0.8700 & 0.8670 & 0.9745& 7.3576 & 2.1834 \\
DAN\cite{zhang2017deep} + Swin-UNet & 0.8288 & 0.9936 & 0.8261 & 0.8375 & 0.9721& 9.9132 & 2.7309 \\
ICT\cite{verma2019interpolation} + Swin-UNet &  0.8621 & 0.9947 & 0.8624 & 0.8632 & 0.9746 & 8.7211 & 2.5562\\
ADVENT\cite{vu2019advent}+ Swin-UNet & 0.8669 & 0.9949 & 0.8688 & 0.8660 & 0.9743 & 7.1383 & 2.2608 \\
UAMT\cite{yu2019uncertainty} + Swin-UNet\cite{wang2022uncertainty} & 0.8701 & 0.9950 & 0.8721 & 0.8697 & 0.9754 &  6.7226 & 2.0975\\
DCN\cite{qiao2018deep} + Swin-UNet  & 0.8608 & 0.9946 & 0.8511 & 0.8724 & 0.9777 & 8.8474 & 2.6705 \\
CPS\cite{chen2021semi} + Swin-UNet & 0.8933 & 0.9957 & 0.8846 & 0.9032 & 0.9821& 5.5661 & 1.6418\\
\hline
MT\cite{tarvainen2017mean} + UNet &  0.8781 & 0.9949 & 0.8836 & 0.8735 & 0.9690 & 10.9691 & 3.3246 \\
DAN\cite{zhang2017deep} + UNet & 0.8766 & 0.9948 & 0.8814 & 0.8727 & 0.9700 & 8.6977 & 2.4750\\
ICT\cite{verma2019interpolation} + UNet & 0.8879 & 0.9953 &  0.8996 & 0.8779 & 0.9696 & 6.7011 & 1.9696  \\
ADVENT\cite{vu2019advent}+ UNet & 0.8777 & 0.9949 & 0.8877 & 0.8703 & 0.9674 & 11.0979 & 2.9367 \\
UAMT\cite{yu2019uncertainty} + UNet & 0.8798 & 0.9949 & 0.8778 & 0.8821 & 0.9726& 10.2134 & 3.1926 \\
DCN\cite{qiao2018deep} + UNet & 0.8831 & 0.9952 & 0.8897 & 0.8785 & 0.9706 & 8.6978 & 2.7026 \\
CPS\cite{chen2021semi} + UNet & 0.8933 & 0.9956 & 0.8965 & 0.8912 & 0.9749 &7.8319 & 2.2767\\
\hline
Semi-Mamba-UNet & {\bf 0.9114 } &{\bf 0.9964} & {\bf0.9088} & {\bf0.9146} & {\bf0.9821} & {\bf 3.9124} & {\bf 1.1698} \\
\hline
\end{tabular}
\label{tablebaseline222}
\end{table*}

\begin{table*}[htbp]
\footnotesize
\caption{Direct comparison of semi-supervised frameworks on MRI prostate test set when eight cases of 50 data were assumed as labelled data.}
\centering
\begin{tabular}{c|ccccc|cc}
\hline
SSL Framework+Network & Dice$\uparrow$  & Acc$\uparrow$ & Pre$\uparrow$ & Sen$\uparrow$ & Spe$\uparrow$ & HD$\downarrow$ & ASD$\downarrow$  \\
\hline
MT\cite{tarvainen2017mean} + Swin-UNet & 0.7249 & 0.9889 & 0.8097 & 0.6562 & 0.9965 & 12.6295 & 3.3809 \\
DAN\cite{zhang2017deep} + Swin-UNet &  0.6973 & 0.9877 & 0.7737 & 0.6346 & 0.9958 & 13.6035 & 3.6825\\
ICT\cite{verma2019interpolation} + Swin-UNet & 0.6925 & 0.9876 & 0.7707 & 0.6287 & 0.9957& 13.3477 & 3.6555\\
ADVENT\cite{vu2019advent}+ Swin-UNet  & 0.6870 & 0.9875 & 0.7784 & 0.6149 & 0.9960 & 24.3831 & 5.6376 \\
UAMT\cite{yu2019uncertainty} + Swin-UNet\cite{wang2022uncertainty} & 0.6987 & 0.9878 & 0.7803 & 0.6325 & 0.9959 & 13.4233 & {\bf 3.1231}  \\
DCN\cite{qiao2018deep} + Swin-UNet & 0.7341 & 0.9891 & 0.8087 & 0.6721 & 0.9964 & 21.8991 & 4.1590\\
CPS\cite{chen2021semi} + Swin-UNet & {\bf 0.7635} & 0.9895 & 0.7694 & {\bf 0.7577} & 0.9948 & 10.5098 & {\bf 2.6318}  \\
\hline
MT\cite{tarvainen2017mean} + UNet  & 0.5940 & 0.9821 & 0.6021 & 0.5862 & 0.9912 & 35.1110 & 9.3841 \\
DAN\cite{zhang2017deep} + UNet & 0.5862 & 0.9827 & 0.6292 & 0.5486 & 0.9926 & 39.9249 & 11.4807 \\
ICT\cite{verma2019interpolation} + UNet & 0.7610 & 0.9902 &\bf 0.8301 & 0.7025 & 0.9967 & 18.7682 & 4.9913 \\
ADVENT\cite{vu2019advent}+ UNet & 0.6098 & 0.9860 & 0.8033 & 0.4914 & 0.9973 & 30.9931 & 8.8792\\
UAMT\cite{yu2019uncertainty} + UNet & 0.5935 & 0.9854 & 0.7780 & 0.4798 & 0.9969 & 64.5963 & 19.4225 \\
DCN\cite{qiao2018deep} + UNet & 0.6769 & 0.9871 & 0.7643 & 0.6075 & 0.9957 &  60.3634 & 15.9329 \\
CPS\cite{chen2021semi} + UNet &  0.7067 & 0.9888 & {\bf 0.8514} & 0.6040 & 0.9976 & 19.4960 & 4.5235 \\
\hline
Semi-Mamba-UNet & \underline{ 0.7627} & {\bf 0.9901} & {\underline {0.8172}} & \underline{0.7151} & {\bf 0.9965} & {\bf 10.4211} & \underline{3.3760 } \\
\hline
\end{tabular}
\label{prostatetablebaseline111}
\end{table*}

\begin{table*}[htbp]
\footnotesize
\caption{Direct comparison of semi-supervised frameworks on MRI prostate test set when 12 cases of 50 data were assumed as labelled data.}
\centering
\begin{tabular}{c|ccccc|cc}
\hline
SSL Framework+Network & Dice$\uparrow$  & Acc$\uparrow$ & Pre$\uparrow$ & Sen$\uparrow$ & Spe$\uparrow$ & HD$\downarrow$ & ASD$\downarrow$  \\
\hline
MT\cite{tarvainen2017mean} + Swin-UNet & 0.7429 & 0.9897 & 0.8367 & 0.6680 & 0.9970 & 10.0029 & 2.3684  \\
DAN\cite{zhang2017deep} + Swin-UNet & 0.7428 & 0.9891 & 0.7864 & 0.7037 & 0.9956 & 10.3433 & 2.8220 \\
ICT\cite{verma2019interpolation} + Swin-UNet & 0.7671 & 0.9908 & 0.8857 & 0.6764 & {\bf0.9980} & 8.0819 & 2.0044 \\
ADVENT\cite{vu2019advent}+ Swin-UNet  & 0.6971 & 0.9885 & 0.8445 & 0.5935 & 0.9975 & 10.4619 & 3.0904 \\
UAMT\cite{yu2019uncertainty} + Swin-UNet\cite{wang2022uncertainty}  & 0.7518 & 0.9899 & 0.8262 & 0.6897 & 0.9967 & 9.9750 & 2.8523 \\
DCN\cite{qiao2018deep} + Swin-UNet & 0.6832 & 0.9876 & 0.7968 & 0.5980 & 0.9965 & 12.8645 & 2.7548\\
CPS\cite{chen2021semi} + Swin-UNet & 0.8008 & 0.9917 & 0.8652 & 0.7453 & 0.9974 & 9.4535 & 2.1138 \\
\hline
MT\cite{tarvainen2017mean} + UNet  & 0.6623 & 0.9870 & 0.7863 & 0.5721 & 0.9965 & 47.1670 & 10.4780\\
DAN \cite{zhang2017deep} + UNet  & 0.6491 & 0.9877 & 0.8926 & 0.5100 & 0.9986 & 18.6358 & 4.1067\\
ICT\cite{verma2019interpolation} + UNet  & 0.7211 & 0.9897 & {\bf 0.9032 } & 0.6001 & 0.9985 & 12.9646 & 4.3894  \\
ADVENT\cite{vu2019advent}+ UNet & 0.6657 & 0.9877 & 0.8491 & 0.5474 & 0.9977 &18.6658 & 4.0472\\
UAMT\cite{yu2019uncertainty} + UNet & 0.6775 & 0.9867 & 0.7397 & 0.6249 & 0.9950 & 62.7224 & 15.1459 \\
DCN\cite{qiao2018deep} + UNet  & 0.6673 & 0.9877 & 0.8430 & 0.5522 & 0.9977 &15.7993 & 4.2646\\
CPS\cite{chen2021semi} + UNet & 0.7233 & 0.9894 & 0.8701 & 0.6188 & 0.9979  & 21.3932 & 5.8624\\
\hline
Semi-Mamba-UNet & {\bf 0.8216} & {\bf0.9926 } & 0.8842 & {\bf 0.7672} & \underline{0.9977}  & {\bf9.9395} & {\bf2.4943} \\
\hline
\end{tabular}
\label{prostatetablebaseline222}
\end{table*}
\textbf{Evaluation Metrics:} Comprehensive evaluation metrics were employed to assess the performance of Semi-Mamba-UNet against other SSL baseline methods. The similarity measures included Dice, accuracy (Acc), precision (Pre), sensitivity (Sen), and specificity (Spe).

\begin{align}
\text{Dice} & = \frac{2 TP}{2 TP + FP + FN} \\
\text{Accuracy} & = \frac{TP + TN}{TP + TN + FP + FN} \\
\text{Precision} & = \frac{TP}{TP + FP} \\
\text{Sensitivity} & = \frac{TP}{TP + FN} \\
\text{Specificity} & = \frac{TN}{TN + FP}
\end{align}

where TP, FP, TN, and FN represent the numbers of true positives, false positives, true negatives, and false negatives, respectively, for each pixel. These metrics provide a comprehensive assessment of the network performance across various aspects; that is, the higher, the better, as denoted by $\uparrow$.

In addition, the difference measures, where lower values are preferable$\ (downarrow$, consisted of the 95\% Hausdorff distance (HD) and average surface distance (ASD). Given that one of the datasets involves four-class segmentation tasks, the mean values of these metrics across all classes are reported.

{\bf Qualitative Results:} Figures~\ref{fig:3} and \ref{fig:7}illustrate the three randomly selected sample raw MRI cardiac scans, ground truth, and corresponding predictions of all SSL baseline frameworks with several types of UNet, including Semi-Mamba-UNet, under different data situations when three and seven cases of the total of 100 cases in the training set were labelled data for training, respectively. Figures \ref{fig:8} and \ref{fig:12} illustrate the three randomly selected sample raw MR prostate scans, ground truth, and corresponding predictions of all SSL baseline frameworks with several types of UNet, including Semi-Mamba-UNet, under different data situations when eight and 12 cases of the total of 50 cases in the training set were labelled data for training, respectively.

\textbf{Quantitative Results:} The performance of Semi-Mamba-UNet in direct comparison with other SSL methods on the MRI cardiac dataset is quantitatively detailed in Tables~\ref{tablebaseline111} and~\ref{tablebaseline222}, encompassing both similarity and difference measures under the condition when three and seven cases of 100 cases of training data were labelled. Tables \ref{prostatetablebaseline111} and \ref{prostatetablebaseline222} report the results on the MR prostate test set under two different data situations. In all of the above tables, the highest-performing metrics are highlighted in \textbf{bold}, and the second-best of Semi-Mamba-UNet is \underline{underlined}.

\begin{table*}[t]
\footnotesize
\caption{Ablation studies on segmentation backbone network with the same SSL framework on MRI cardiac segmentation test set.}
\begin{center}
\centering
\begin{tabular}{ c|c|ccccc|cc }
\hline
Labelled Cases  &  Network & Dice$\uparrow$  & Acc$\uparrow$ & Pre$\uparrow$ & Sen$\uparrow$ & Spe$\uparrow$ & HD$\downarrow$ & ASD$\downarrow$ \\
 \hline
3 & 2 $\times$ Swin-UNet &0.7878 & 0.9918 & 0.8066 & 0.7795 & {\bf 0.9577} & 9.0240 & 2.3592 \\
3 & 2 $\times$ Mamba-UNet &0.8025 & 0.9924 & 0.8623 & 0.7558 & 0.9379 & 7.3952 & 2.1257\\
3 & UNet + Swin-UNet & 0.8292 & 0.9933 & 0.8591 & {\bf 0.8052} & 0.9557 & {\bf 5.7014} & 1.7237\\
\hline
3 & UNet + Mamba-UNet & {\bf 0.8386} & {\bf 0.9936} &{\bf 0.8861} & 0.7992 & 0.9483 & \underline{6.2139} & {\bf 1.6406} \\
 \hline \hline
7 & 2 $\times$ Swin-UNet & 0.8899 & 0.9955 & 0.8784 & 0.9031 & 0.9823& 5.9222&1.6960 \\
7 & 2 $\times$ Mamba-UNet & 0.9006 & 0.9959 & 0.8913 & 0.9109 & {\bf 0.9826} & 6.7631 & 1.8349 \\
7 & UNet + Swin-UNet & 0.9105 & 0.9963 & 0.9057 & {\bf 0.9161} &  {\bf 0.9826} & 5.4172 & 1.4506\\
\hline
7 & UNet + Mamba-UNet & {\bf 0.9114} & {\bf 0.9964} & {\bf 0.9088} & \underline{0.9146} & 0.9821 & {\bf 3.9124} & {\bf 1.1698} \\
 \hline
\end{tabular}
\label{tab:ablationcomparison}
\end{center}
\end{table*}

\begin{table*}[t]
\footnotesize
\caption{Ablation studies on segmentation backbone network with the same SSL framework on MRI prostate segmentation test set.}
\begin{center}
\centering
\begin{tabular}{ c|c|ccccc|cc }
\hline
Labelled Cases &  Network & Dice$\uparrow$  & Acc$\uparrow$ & Pre$\uparrow$ & Sen$\uparrow$ & Spe$\uparrow$ & HD$\downarrow$ & ASD$\downarrow$ \\
 \hline
8 & 2 $\times$ Swin-UNet & 0.6705 & 0.9877 & 0.8285 & 0.5631 & 0.9973 &25.5297 & 8.9403\\
8 & 2 $\times$ Mamba-UNet & 0.7115 & 0.9889 & {\bf 0.8466} & 0.6136 & {\bf 0.9975} & 11.0528 & 4.7265\\
8 & UNet + Swin-UNet & 0.7600 & {\bf 0.9903} & 0.8468 & 0.6893 & 0.9972 &17.6143 & 5.0325\\
\hline
8 & UNet + Mamba-UNet &{\bf 0.7627} & \underline{0.9901} &  0.8172 & {\bf 0.7151} &  0.9965 & {\bf 10.4211} & {\bf 3.3760 }\\
 \hline \hline
12 & 2 $\times$ Swin-UNet &  0.7362 & 0.9895 & 0.8394 & 0.6556 & 0.9971&13.2056 & 2.7923 \\
12 & 2 $\times$ Mamba-UNet & 0.7402 & 0.9896 & 0.8370 & 0.6635 & 0.9971 & 12.8540 & 2.7876\\
12 & UNet + Swin-UNet &0.8146 & 0.9924 & 0.8907 & 0.7504 & 0.9979 &10.4208 & 2.6860\\
\hline
12 & UNet + Mamba-UNet & {\bf 0.8216} & {\bf0.9926 } & 0.8842 & {\bf 0.7672} & \underline{0.9977}  & {\bf9.9395} & {\bf2.4943} \\
 \hline
\end{tabular}
\label{tab:ablationcomparison2}
\end{center}
\end{table*}

\begin{figure*}
\centering  
\includegraphics[width=\linewidth]{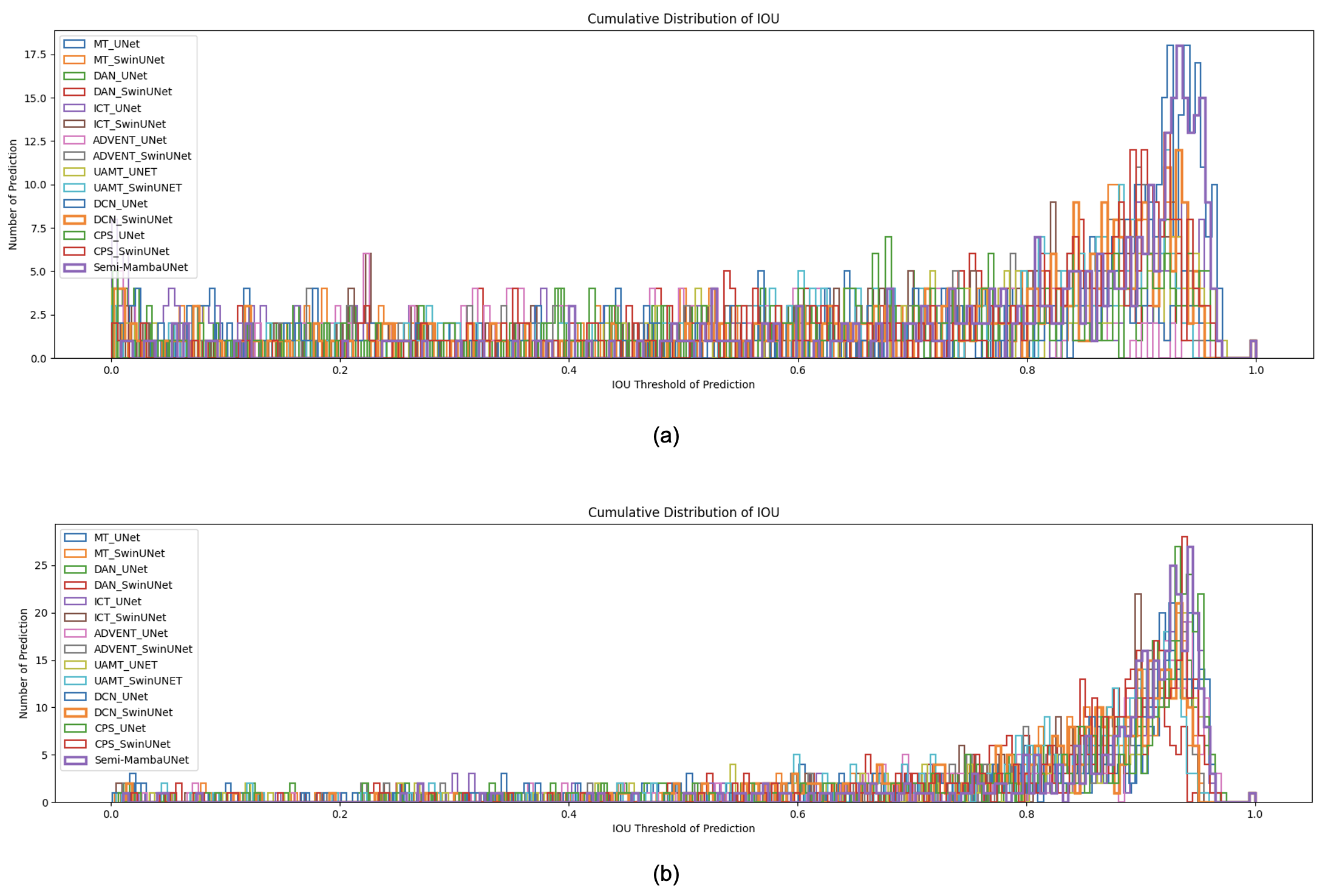}  
\caption{Distribution of segmentation results on MRI cardiac test set according to IoU: (a) when three of 100 cases of data were labelled data and (b) when seven of 100 cases of data were labelled data. }
\label{fig:histogram111}  
\end{figure*}

\begin{figure*}
\centering  
\includegraphics[width=\linewidth]{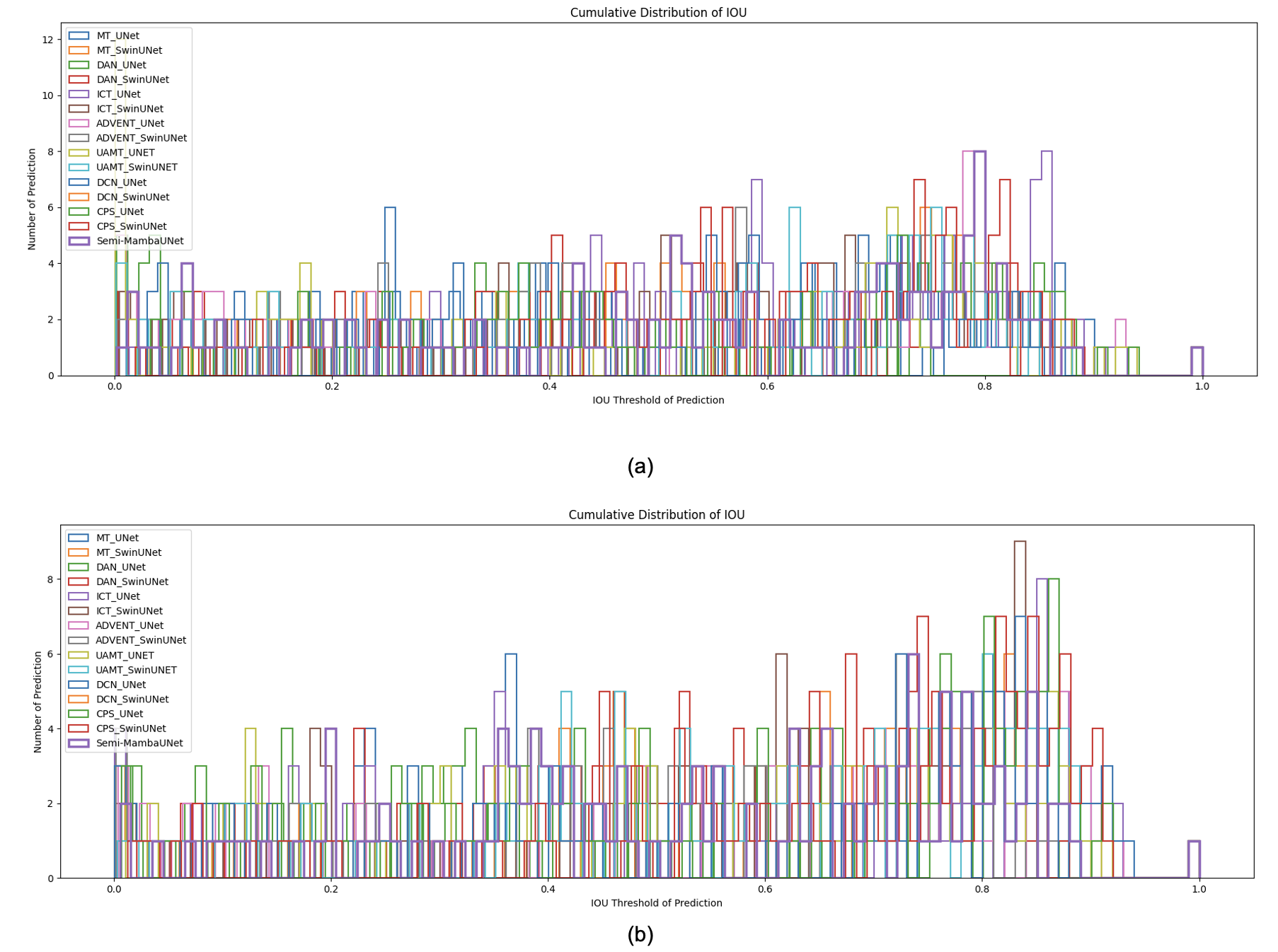}  
\caption{Distribution of segmentation results on MRI prostate test set according to IoU: (a) when eight of 50 cases of data were labelled data and (b) when 12 of 50 cases were labelled data. }
\label{fig:histogram222}  
\end{figure*}

\textbf{Evaluation on Each Image of Test Set:} In addition to reporting the mean values of the evaluation metrics in Tables \ref{tablebaseline111}, \ref{tablebaseline222}, \ref{prostatetablebaseline111}, and \ref{prostatetablebaseline222}, we also present the segmentation results on each image on the test set. Two histogram figures for each dataset are presented in Figures~\ref{fig:histogram111} and \ref{fig:histogram222} with subfigures (a) and (b), respectively, where the x-axis is the IoU and the y-axis is the number of predicted images on the test set with the corresponding IoU score, demonstrating that Semi-Mamba-UNet was more likely to predict images with high IoU scores than the other baseline methods.

\begin{figure*}
\centering  
\includegraphics[width=\linewidth]{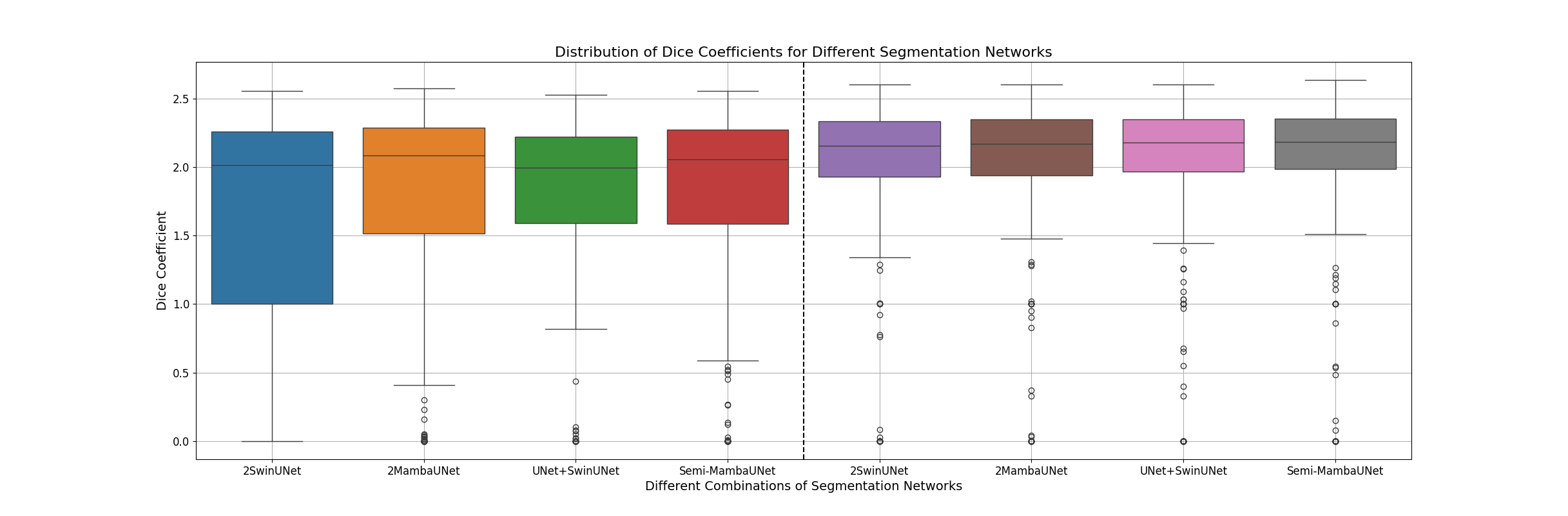}  
\caption{Box plot of segmentation results on test set under two different data situations according to the Dice coefficient in the $tablation study. $-tests were conducted and the $p$-value was $< 0.05$. }
\label{fig:box}  
\end{figure*}

\textbf{Ablation Study:} To evaluate the different combinations of various segmentation networks within our proposed cross-supervised semi-supervised learning strategy further, ablation studies were conducted and the results are presented in Tables \ref{tab:ablationcomparison} and \ref{tab:ablationcomparison2}. CNN-, ViT-, and Mamba-based UNet were selected with different combinations, and the studies were conducted under different data situations on the two datasets. Figure \ref{fig:box} presents a visual comparison of the different combinations of segmentation network settings using a box plot, where the x-axis indicates the network setting and the y-axis indicates the distribution of predictions with the Dice coefficient. A T-test was conducted, and the $P$-value between Semi-Mamba-UNet and the other methods was always lower than 0.05, demonstrating a statistically significant difference in performance. These results indicate the proper setting of Semi-Mamba-UNet, that is, a different segmentation network as a network perturbation for SSL. The differences in the network are considered as a distribution of multi-view learning, and both networks can be beneficial to one another owing to the strength of the local information learning and global feature learning.

\section{Conclusion}
In this study, we investigated the integration of Visual Mamba within the UNet architecture in a semi-supervised manner for medical image segmentation. An advanced SSL strategy combining pixel-level cross-supervision and contrastive learning was proposed to harness the full potential of Visual Mamba. Our extensive experimental evaluations demonstrated the effectiveness of Semi-Mamba-UNet. In the future, we aim to extend our research to encompass volumetric data segmentation and further refine our methods within the scope of limited supervised learning scenarios, continuing to leverage the unique capabilities of Visual Mamba.

C.M.: conceptualisation, data curation, formal analysis, investigation, methodology, resources, software, validation, visualisation, writing—original draft, writing—review and editing. Z.W.: conceptualisation, formal analysis, investigation, methodology, project administration, supervision, validation, writing—original draft, review, and editing. All authors have read and agreed to the published version of the manuscript.

\bibliographystyle{elsarticle-num} 
\bibliography{cas-refs}

\end{document}